# Simulating the time series of a selected gene expression profile in an agent-based tumor model

**Yuri Mansury [1] and Thomas S. Deisboeck [1,2,3*]**

[1] Complex Biosystems Modeling Laboratory, Harvard-MIT (HST) Athinoula A. Martinos Center for Biomedical Imaging, HST-Biomedical Engineering Center, Massachusetts Institute of Technology, Cambridge, MA 02139; [2] Molecular Neuro-Oncology Laboratory, Massachusetts General Hospital, Harvard Medical School, Charlestown, MA 02129; [3] Division of Engineering and Applied Sciences, Harvard University, Cambridge, MA 02138.

**Running Title:** Gene expression profiles in an agent-based tumor model

**\*Corresponding Author:**

Thomas S. Deisboeck, M.D.
Complex Biosystems Modeling Laboratory
Harvard-MIT (HST) Athinoula A. Martinos Center for Biomedical Imaging
HST-Biomedical Engineering Center, Bldg. 16-352
Massachusetts Institute of Technology
77 Massachusetts Avenue
Cambridge, MA 02139
Tel: (617)-452-2226
Fax: (617)-253-2514
Email: deisboec@helix.mgh.harvard.edu

# ABSTRACT

To elucidate the role of environmental conditions in molecular-level dynamics and to study their impact on macroscopic brain tumor growth patterns, the expression of the genes Tenascin C and PCNA in a 2D agent-based model for the migratory trait is calibrated using experimental data from the literature, while the expression of these genes for the proliferative trait is obtained as the model output. Numerical results confirm that the gene expression of Tenascin C is consistently higher in the migratory glioma cell phenotype and show that the expression of PCNA is consistently higher among proliferating tumor cells. Intriguingly, the time series of the tumor cells' gene expression exhibit a sudden change in behavior during the invasion of the tumor into a nutrient-abundant region, showing a robust positive correlation between the expression of Tenascin C and the tumor's diameter, yet a strong negative correlation between the expression of PCNA and the diameter. These molecular-level dynamics correspond to the emergence of a structural asymmetry in the form of a bulging tumor rim in the nutrient-abundant region. Within such geographic regions, the simulated time series thus supports the critical role of the migratory cell phenotype during both the tumor system's overall macroscopic expansion and the evolvement of regional growth patterns, particularly in the later stages. Furthermore, detrended fluctuation analysis (DFA) suggests that for prediction purposes, the simulated gene expression profiles of Tenascin C and PCNA that were determined separately for the migrating and proliferating phenotypes exhibit lesser predictability than those of the phenotypic mixture combining all viable tumor cells typically found in clinical biopsies. Finally, dividing the tumor into distinct geographic regions of interest (ROI) reveals that the gene expression profile of tumor cells residing in the quadrant close to the nutrient-abundant region is representative for the entire tumor whereas the expression profile of tumor cells residing in the geographically opposite ROI is not. Potential implications of these modeling results for experimental and clinical cancer research are discussed.

# 1. INTRODUCTION

Despite all efforts, the outcome of patients suffering from highly malignant neuroepithelial brain tumors remains dismal. Treatment difficulties arise from the main characteristics of these glial tumors such as rapid growth, extensive tissue invasion, and cell heterogeneity. Histological and immunocytochemistry methods [1] and more recently gene expression arrays [2] reveal that the behavior of these tumor cells appears to exhibit a *dichotomy* between proliferation and migration. That is, the same cell can either proliferate or migrate, yet not both at the same time, which should greatly influence the spatio-temporal dynamics of the malignant biosystem. Therefore, analyzing the impact of environmental factors on gene expression changes, which in turn are expected to regulate the phenotypic 'switch' from proliferation to migration, is key to a better understanding of tumor behavior and necessary in order to develop predictive modeling platforms for clinical applications.

As a first step, we employ here a previously developed 2D agent-based model of a virtual multicellular brain tumor system [3,4], which has been augmented to take into account explicitly molecular level dynamics. Our modeling platform is particular suitable for this purpose due to its 'monoclonal' nature, i.e., it focuses on the behavior of a population of homogeneous cells. To render our model tractable, the focus of our very simplified *gene network* is centered on the behavior of two genes only, namely *Tenascin C* and *PCNA*, which are chosen based on their reported roles during proliferation and migration of glioma cells. More specifically, Tenascin C is a an extracellular matrix glycoprotein known to be over-expressed in human gliomas [5,6]. Further, it has been demonstrated *in vitro* that migration is stimulated when human SF-767 glioma cells are placed on Tenascin C [7]. In contrast, the gene expression of proliferating cell nuclear antigen (PCNA) markedly rises during G1/S phase, thus serves as an indicator of proliferation activities [8]. High levels of PCNA have been shown to correlate with histological grade and poor prognosis in glioma patients [9,10].

By calibrating the expression of both genes using experimental data for the migratory phenotype while generating the gene expression for proliferating and quiescent cells as the model output, numerical results of the model confirm that among the migratory phenotype the expression of Tenascin C is indeed consistently higher, while they reveal the reverse for the proliferating tumor cells, which exhibit consistently higher expression of gene PCNA. In our particular modeling setup, this *genetic switch*, determining the cellular *phenotype*, can be explained by the tendency for migrating cells to encounter higher level of hypoxia than their proliferating peers. We then show that these molecular level dynamics can be directly linked to the tumor diameter, i.e., to the structural characteristic at the multicellular level. More specifically, in the presence of abundant nutrients, the gene expression profiles indicate a breaking point in the time series after which we find robust (*i*) positive correlations between tumor diameter and the expression of Tenascin C, and (*ii*) negative correlations between the former and PCNA expression, both as statistical averages for the entire pool of viable tumor cells. Furthermore, during tumor invasion towards the geographic region of abundant nutrients, these molecular dynamics are accompanied by the emergence of a structural asymmetry in the rim of the growing tumor. The numerical results therefore support the notion that in the presence of these distinct microenvironmental conditions the migratory phenotype is crucial for the tumor system's macrocopic expansion and its regional structural patterns. In addition, detrended fluctuation analysis (DFA) suggests that the gene expression profile of a phenotypically mixed cell population has a higher predictive value than that of a virtual tumor specimen that is divided into separate migrating and proliferating phenotypic subsets. Lastly, dividing the tumor into distinct geographic regions of interest (ROI) reveals that the quadrant close to the nutrient-abundant region, which also harbors the macroscopically conspicuous rim area, is representative for the entire tumor in terms of DFA and molecular-structural correlation results whereas the geographically opposite ROI is not.

In the next section we will first describe the mathematical model in more detail followed by the specific oncology concept implemented here.

# 2. MATHEMATICAL MODEL

## A. Algorithm

The complete details of our agent-based model are described elsewhere [3,4,11]. Briefly:

- *Microenvironment:* environmental variables in our model consist of nutrient levels, toxic metabolites, and mechanical confinements. First, nutrient sources evolve according to the following equation of motion: $\Delta \phi_j = g_\phi \phi_{t-1,j} \Delta t + D_\phi \nabla(\phi_{t-1}) \Delta t - r_\phi \Delta l_{t-1,j}$, with $g_\phi$ representing the constant rate of nutrient replenishment, $D_\phi$ the nutrient diffusion coefficient, and $r_\phi$ the nutrient depletion rate. Similarly, the levels of toxic metabolites are also updated using a linear difference equation: $\Delta \tau_j = D_\tau \nabla(\tau_{t-1}) \Delta t + r_\tau \Delta l_{t-1,j}$. Finally, the dynamic evolution of mechanical confinements represents the 'adaptive grid lattice' property of the host matrix, such that locations that have been invaded by tumor cells in the past will subsequently experience a decline in 'tissue consistency' and thus in mechanical confinement: $p_{t,j} = p_{t-1,j} - r_p \, l_{t-1,j}$.

- *Cell behavior:* the aforementioned dichotomy between growth and invasion [1,2] in our model is captured by alternating sequences of "proliferation-migration-proliferation" behavior, which in turn drive the formation of distinct spatiotemporal patterns. Specifically, the onset of both proliferation and migration must satisfy the dual threshold levels of nutrients, ($\phi_L, \phi_U$), and toxicity, ($\tau_L, \tau_U$) such that a tumor cell in location $j$ proliferates if $\phi_j > \phi_U$ and $\tau_j < \tau_L$, migrates if $\phi_L < \phi_j < \phi_U$ and $\tau_L < \tau_j < \tau_U$, else the cell dies (representing apoptosis) or enters a reversible, quiescent state. For tumor cells that fail to fulfill the proliferation criteria but are eligible to migrate, they assign a value $T_j$ to a location $j$ based on the following $F : \Omega \to \Re$ function: $F_j = \Psi \, L_j + (1 - \Psi) \xi_j$, where $\Omega$ is the set of locations that are adjacent to that tumor cell's current position, $L_j$ is the error-free component (the explicit specification of $L_j$ is derived in [3,4]), and $\xi_j$ is an error term distributed as $N(0, \sigma^2)$. The parameter $\Psi \in [0,1]$ represents the *precision* of chemotactic local search such that the closer $\Psi$ to unity, then the more accurate the evaluation of location $j$ would be. Based on a "least resistance, least toxicity and highest attraction paradigm" the migrating tumor cell then moves toward the most permissive location $j$ in the choice set $\Omega$, i.e., arg max $_{j \in \Omega} T_j (L_j)$.

- *Lattice setup:* Based on this mathematical description, our simulation platform is initially setup as follows. As in previous works, our spatial backdrop is a 200 x 200 torroidal lattice, representing a virtual section of brain tissue. The initial spatial distributions of nutrients and mechanical confinements are detailed elsewhere [3,4]. In brief, there are two replenished primary nutrient sources, representing cerebral blood vessels, located at the center and at the northeast (NE) quadrant of the lattice. To establish a strong chemotactic gradient, the peak of the second source at the NE quadrant is set up to be 5-times larger than that of the first source at the lattice center. Outside of these two sources, there are much smaller sources of non-replenished ('interstitial') nutrient substrates that are randomly-uniformly distributed. In contrast, we assume that the distribution of mechanical confinements follows a uniform distribution everywhere, except inside of a small 'crater' or void at the center of the grid lattice, adjacent to the first primary nutrient source. As in previous works, within the radius of the two primary nutrient sources, the initial levels of toxicity are assumed to be zero. As the tumor grows, the level of 'toxic' metabolites rises representing the reduction in pH and $pO_2$ in densely populated areas as well as the byproducts released from dying tumor cells. However, and unlike in previous works, outside of the radius of the two primary nutrient sources, here we introduce an initial distribution of toxic metabolites, representing here in addition the effect of 'hypoxia' that are already in the system from the beginning. These initial levels of hypoxia are generated based on a uniform probability distribution function. At the beginning of every simulation, the grid lattice is seeded with 10 tumor cells that are placed in the 'crater' at the center of the grid lattice. To facilitate initial invasion into the surrounding matrix, it is assumed that at t = 0 the crater exhibits low mechanical stresses and contains no nutrients. At t = 1, both nutrients and toxic metabolites start to diffuse into the crater due to declining mechanical constraints brought about by ('tissue architecture-disrupting') tumor cell invasion. The simulation terminates when the first tumor cell has reached the peak of the second primary nutrient source.

The following section details the experimental basis for the gene regulatory network operating at the molecular modeling level.

### B. Experimental data and 'gene expression-environment interaction' concept

Using complementary DNA (cDNA) NIH microarrays, Mariani et al. [2] reported that the gene expression of Tenascin C is 4-fold *up*-regulated in human G112 glioma cells that were cultured on an extracellular matrix (ECM) derived from glioma cells. Moreover, compared to the control experiment, the authors found that these glioma cells exhibit a significantly accelerated migration rate of more than 250 μm/day. Concomitantly, that same migratory-stimulated glioma cells show a 2-3.5-fold *down*-regulation of PCNA. Taken together, these findings finding suggest that the migratory phenotype in these glioma cells is associated with both up-regulation of Tenascin C and down-regulation of PCNA.

## TABLE I.

**Table I** lists the cDNA NIH microarray gene expression values for both genes reported in Mariani et al. [2]. Our approach is to first, calibrate the initial levels of Tenascin C and PCNA for migrating tumor cells in our agent-based model according to these data. The expression of these genes for proliferating (and quiescent) tumor cells is then obtained *endogenously*, i.e., as the output of our model. The impact of the microenvironmental conditions on gene expression is computed as follows.

- **Tenascin C** has been shown to be strongly expressed in human GaMG glioblastoma cells that were cultured in filtrated medium containing 10 percent fetal calf serum, yet showed a marked decrease when cultured in serum-free conditions [12]. In addition, with the use of serial analysis of gene expression (SAGE), Lal et al. [13] found that under hypoxic conditions, the gene encoding Tenascin C is induced in human D247-MG glioblastoma cells. Together, these findings suggest that an *increase* in both the onsite levels of nutrients ('serum') and toxicity ('hypoxia') leads to an *increase* of Tenascin gene expression (*gTenascin*). In our model, for each cell in the grid lattice we transform the levels of nutrients, $\phi_j$, toxicity, $\tau_j$, into the scalars $\hat{\phi}_j = \exp[\varphi_j /(\varphi_j + a_\phi)]$ and $\hat{\tau}_j = \exp[\tau_j /(\tau_j + a_\tau)]$, where $a_\phi$ and $a_\tau$ are parameter constants. Subsequently,

*gTenascin* is computed as a positive, multiplicative function of both nutrients and toxic metabolites, $gTenascin = b_T \ \hat{\phi}_j \ \hat{\tau}_j$. The transformation procedure and the parameter $b_T$ ensure that $gTenascin \in [0,10]$ and that its initial level equals 4.88 for migrating tumor cells (see **Table I**). Note that due to the absence of specific comparative experimental data, this formulation assigns equal, positive *weights* to both environmental variables.

- Folkman [14] had argued that for a tumor to exceed 2-3 mm in diameter, it requires angiogenesis as it relies on perfusion for further growth. Indeed, brain tumor cell proliferation is especially marked in perfused tumor areas [15]. As has been shown for osteoblasts, cell proliferation and thus **PCNA** expression should therefore decrease under hypoxic conditions [16], such as in locations farther away from blood vessels, due to a lack of oxygen related to its diffusion limit of 100-200 μm [17]. Yet, like Tenascin C, the PCNA gene (promoter) also appears to be serum-responsive [18]. These data together suggest that an *increase* in nutrients and a *decrease* in hypoxia, i.e. toxicity lead to an *increase* of PCNA gene expression (*gPCNA*). Accordingly, we compute $gPCNA = b_P \ \hat{\phi}_j \ / \hat{\tau}_j$, where $b_P$ is selected such that $gPCNA \in [0,10]$ and that the initial level of *gPCNA* equals 0.285 for migrating tumor cells (**Table I**).

In the following section we will briefly describe the numerical results of the model.

## 3. RESULTS

The typical simulated time series of (a) Tenascin C and (b) PCNA for proliferating and migrating tumor cells from a typical run of our model is shown in **Fig. 1**.

## FIG. 1

The gene expression of Tenascin C for migrating cells is always higher than that for proliferating cells. In contrast, the gene expression of PCNA for proliferating cells is consistently higher than that for migrating cells. These findings suggest that our algorithm operates as a *genetic switch* that simultaneously down-regulates *gTenascin* and up-regulates *gPCNA* for proliferating cells relative to those of their migrating peers. Not shown in **Fig. 1** is the gene profile for quiescent tumor cells, whose expression of both *gTenascin* ($3.59 \pm 0.27$) and *gPCNA* ($0.16 \pm 0.01$) tend to be the lowest relative to migrating and proliferating cells.

Another prominent property of the time series is the sharp trend change at a particular *breakpoint*, indicating non-stationary stochastic processes. **Figure 2** reveals that as the tumor system begins to invade the second nutrient source at the NE region of the grid lattice, the average *gTenascin* across all viable cells experiences a steep rise concomitant with a sudden decline in *gPCNA* (for the simulation shown, at approximately $t = 308$). **Figure 3** indicates that this alteration in the gene expression profile is accompanied by the emergence of a structural asymmetry, i.e., a bulging area on the tumor surface adjacent to the peak of the second primary nutrient source (located in the *top right* corner, i.e., the NE region of the lattice).

**FIG. 2**

**FIG. 3**

The time series depicted both in **Figs. 1** and **2** exhibit both fluctuating "roughness" that could be described by a power law relationship, as well as non stationarity due to the changing means. To characterize the dynamic structure of the time series, we employ *detrended fluctuation analysis (DFA)* that has been previously developed to examine non-stationary signals [19]. Furthermore, DFA has been shown to be a robust method in detecting long-range correlations in heterogeneous DNA sequences [20,21]. Briefly, the DFA method entails the following steps. First, divide each time series of gene expression $\{g_i\}$ of length T into T/$\omega$ nonoverlapping windows, each containing $\omega$ data points. Next, in each window of size $\omega$, "detrend" the time series by regressing gene expression against a time trend and an intercept term, $\widetilde{g}_{i,\omega,t} = a + bt$, then compute the minimum sum-of-squared residuals (*SSR*) for each window equals

$$SSR_{i,\omega,p} = \sum_{t=p}^{p+\omega} (g_{i,t} - \widetilde{g}_{i,\omega,t})^2 .$$ Finally, calculate the average $SSR_{i,\omega,p}$ over all windows of

size $\omega$, $\overline{SSR}_{i,\omega} = 1/(T/\omega)\sum_p SSR_{i,\omega,p} .$ If the time series exhibits the statistical properties

of a *random walk* (i.e., no autocorrelations across time), then DFA would yield

$\sqrt{\overline{SSR}_{i,\omega}} \sim \omega^{0.5} .$ In contrast, if there is a long-range power-law autocorrelation, then

$\sqrt{\overline{SSR_{i,\omega}}} \sim \omega^{\alpha}$ with $\alpha \neq 0.5$. The presence of a long-range autocorrelation that is significantly different from 0.5 indicates the potential use of past values to forecast future realizations using for example a linear state-space method [22,23].

## TABLE II.

Columns (1) in **Table II** show the Monte Carlo simulation results from applying DFA to the time series of Tenascin C and PCNA gene expression in distinct regions of the tumor corresponding to (I) the entire tumor, (II) the NE region where the second nutrient source is located, and (III) the opposite, i.e. southwest (SW) region. Regardless of our region of interest (ROI), it can be seen that when DFA is applied separately to proliferating and migrating cells, the magnitude of $\alpha$ for proliferating cells indicates a stochastic process that is closer to a random walk. However, when DFA is applied to *all* viable tumor cells (i.e., including proliferating, migrating, and quiescent cells) for the entire tumor as well as the NE region, we find the magnitudes of $\alpha$ being significantly different from 0.5 for both *gTenascin* and *gPCNA*, indicating long-range power-law autocorrelation over the entire time series that is inherent in a dynamical system far from equilibrium. Thus, if we have the past realizations of aggregate gene expression in a sample taken from the NE ROI, we can compute the long-range autocorrelation $\alpha$ to predict the short-term trajectory of the structural pattern that is likely to emerge for that tumor region. Furthermore, these NE ROI sample is representative of the entire tumor, as evident from the magnitudes of $\alpha$. In contrast, in the SW region, which contains only low-level nutrient sources, the magnitudes of $\alpha$ resemble those that have been generated by a random walk, and thus have little predictive values for both this particular ROI as well as the entire tumor.

While DFA is a useful method to quantify the dynamic structure of a time series, it does not reveal the extent of the *co-movements* between a pair of variables. It turns out that the correlations between the tumor's macroscopic pattern and its molecular-level gene expression depend on both the timing and the geography of the sampling. For the entire time series, the correlation between the tumor diameter, which represents macroscopic structural characteristics, and *gTenascin* is 0.56 ± 0.05, while between the former and *gPCNA* is –0.52 ± 0.03, both are averages for all viable cells. At first, these rather low correlations seem to suggest a weak relationship between structural pattern and molecular-level dynamics. However, since the time series of *gTenascin* and *gPCNA* are both characterized by non-stationarity as indicated by the changing trend at the breakpoint (see **Fig. 2**), it is necessary to split the series at the breakpoint and then compute the correlation coefficient from that point on, separately for each subsample of interest. Let $C_{T_B,T_E}$ represent the correlation coefficient when the series is truncated between the breakpoint, $T_B$, and the end of the simulation, $T_E$. Columns (2) in **Table II** list $C_{T_B,T_E}$ between, on one hand, the expression of genes Tenascin C and PCNA (averaged over all viable tumor cells) and, on the other hand, the tumor diameter (or the tumor radius for the ROIs). For the entire tumor as well as in the NE quadrant of the lattice, the magnitude of the correlations indicates a robust *positive* correlation between average tumor diameter and *gTenascin* (of all viable tumor cells), and at the same time a strong *negative* correlation between the former and *gPCNA*. In both cases we also found that if we compute $C_{T_B,T_E}$ for the proliferating and migrating phenotypes separately, then only the latter exhibit strong correlations between gene expression and the tumor diameter, although both phenotypic subgroups exhibit *lower* correlations than the virtual specimen that contains a mixture of *all* viable tumor cells. These results are robust as indicated by the small standard errors from performing Monte Carlo simulations. Lastly, the magnitudes of $C_{T_B, T_E}$ again indicate that the sample taken from the region close to a high-level nutrient source (in this case from the NE ROI) is representative of the entire tumor. In contrast, the low-nutrient SW ROI exhibits only weak correlations between the tumor diameter and the expression of genes Tenascin C and PCNA.

In the following section we will briefly discuss these results and their potential impact on experimental and clinical oncology.

# 4. DISCUSSION AND CONCLUSIONS

Quantitative gene expression arrays provide novel and important information for experimental and clinical cancer research alike. However, since such arrays typically represent static snapshots taken at a given point in time from a specific specimen taken out of its 3D tissue context, they do not describe the *dynamics* on the molecular level and hence do not allow one to deduce their dynamic correlations with the overall expanding tumor structure the specimen has been taken from. Integrating this relationship between molecular-level gene expression and the macroscopic tumor structure within a dynamic framework therefore represents an important application for novel computational models, which one can begin to calibrate using already available tumor biology data.

For this purpose, here we have incorporated a gene expression module into a previously developed 2D agent-based tumor model which focuses on both the microscopic cellular level *and* the multicellular patterns arising from interactions with distinct environmental conditions. As a first approximation, this simplified genetic profile or 'network module' includes the expression of only two genes, Tenascin C and PCNA, respectively. Since specific expression levels have been reported for the migratory glioma cell phenotype, we calibrated the expression of genes Tenascin C and PCNA in our model for this phenotype only while endogenously obtaining the gene expression levels for the proliferative and quiescent phenotypes as the output of our model.

The results shown in **Fig. 1** confirm that the Tenascin C gene expression experiences an up-regulation while PCNA is down-regulated during cell migration, whereas the PCNA gene expression is *up*-regulated during cell proliferation with a concomitantly *down*-regulated Tenascin C. The first part is in agreement with the cDNA data reported from [2], implemented in the modeling concept. The proliferative part or, more precisely, the gene-network *switch* itself, leading to the change in phenotypic behavior can be explained by examining the environmental variables that govern the expression of Tenascin C and PCNA. Examination of nutrient and toxicity levels during a typical run reveals that as tumor cells begin to invade the second nutrient source, nutrient levels for migrating cells jump to a level 2.5-fold higher than those for proliferating cells, while toxic metabolites, representing in this iteration primarily hypoxic conditions (within our virtual 2D tissue section), reach a level 16-fold higher for migrating cells than those for proliferating cells. It is the simultaneous increases in nutrients and toxic metabolites that cause the breakpoint in the time series of the Tenascin C and PCNA expression (see **Fig. 2**). This intriguing behavior therefore can be explained by the explicit dependencies of gene expression in our model on environmental variables, which correspond to the fact reported in the literature that the expression of genes Tenascin C and PCNA depend positively on nutrient levels, yet only the Tenascin C is positively related to hypoxia levels while PCNA negatively so. However, it is noteworthy that none of the model equations impose the condition that the expression of genes Tenascin and PCNA in the proliferative cell population, which has been obtained as the output of our model, should remain at levels that are as remarkably *stable* over time as seen in **Fig. 1**.

We also found that in the NE region where tumor cells are attracted by high levels of nutrient sources, representing for example the growth of the malignant brain tumor toward branches of the middle cerebral artery, the average tumor diameter exhibits a strong positive correlation with *gTenascin*, yet an equally strong negative correlation with *gPCNA*. Furthermore, our model also shows that this particular molecular-level behavior is accompanied by the emergence of a structural asymmetry, namely a bulging area, at the tumor surface adjacent to the largest nutrient source representing the aforementioned neighboring blood vessel. These computational findings are in agreement with Hoshino and Wilson [24] who stated that as a rule, an increase in brain tumor size is accompanied, amongst others, by a lower growth fraction, GF. As the total tumor cell number increases, and since the authors had defined GF as the proliferating pool cells divided by the total cell population, an additional decrease in PCNA expression as shown in our simulation results would contribute to an even more substantial GF reduction while the tumor diameter increases. Numerical results of our model confirm that *gTenascin* is up-regulated among migrating cells (**Fig. 1**) to a level higher than for proliferating cells.

Thus, in this simplified network, not only is Tenascin C the gene primarily responsible for the tumor's spatial expansion (into the NE quadrant) as measured by the average tumor diameter, but it is also critical in the development of distinct structural patterns. Cautiously extrapolated to the clinical situation, our numerical results suggest that the expression profile of the gene subset associated with migratory cell activity (e.g., Tenascin C and others) should be monitored at least as closely in brain tumors during the later stages as the often routinely assessed cell proliferation markers.

Furthermore, the DFA results indicate that when *all* viable tumor cells are combined, the time series of gene expression exhibit long-range autocorrelation as indicated by the dynamic fractal dimensions, namely $\alpha = 1.32$ for *gTenascin* and $\alpha = 1.06$ for *gPCNA*. In contrast, when DFA is applied *separately* to migrating and to proliferating cells [columns (1) in **Table II**], the computed $\alpha$ suggests dynamical behavior that is much closer to that of a random walk (i.e., with $\alpha = 0.5$). Therefore, actual biopsy results that almost certainly contain all three tumor phenotypes (i.e., proliferating, migrating, and quiescent cells) appear to be more desirable than separate phenotypic subsets for gene expression profiling in generating time series whose past values can be utilized to predict future realizations, at least during a short-term duration. In contrast, predictions that are exclusively based on the gene expression for the proliferative cell phenotype appear to be relatively weak [columns (1) in **Table II**]; hence the gene expression of this phenotype alone is less useful for detecting co-movements with structural macroscopic patterns [columns (2) in **Table II**]. These findings thus suggest that biopsies from the proliferative tumor core are less useful for predicting macroscopic patterns than biopsies from the tumor edge, which should harbor a larger fraction of migratory phenotypes (i.e., contain a more heterogeneous mix of cell phenotypes). Furthermore, the magnitude of the $\alpha$'s also suggests the extent of autocorrelation is greater for *gTenascin* than for *gPCNA*; thus the prediction of *gTenascin* can be based on past values that had been recorded at larger inter-temporal interval than those required for the prediction of *gPCNA* while maintaining the same error level. This points towards potential implications for clinical oncology as it indicates that in order to predict the gene expression dynamics of Tenascin C in a patient's brain tumor, the interval between invasive control biopsies could in fact be kept longer, hence lowering the burden for the patient without jeopardizing the level of predictability. However, in a real medical setting it would be impossible to collect and analyze all tumor cells of a patient, and hence it is noteworthy that we found marked differences in the predictive value of distinct geographic regions within the tumor. Specifically, the ROI NE closely represents the behavior of the entire tumor in terms of its DFA results whereas ROI SW does not. Taken together with the appearance of the surface asymmetry within the NE ROI, one could argue that in order to maximize its predictive value, a particular tumor tissue biopsy has to be prompted by the emergence of and then image-guided towards such conspicuous tumor rim regions.

It is important to realize the potential shortcomings of our approach. First, the precise nature of the gene regulatory network involving *gTenascin* and *gPCNA* is in large parts still unknown. However, even if more experimental information should become available in the future, it remains to be seen whether a model that incorporates the full complexity of the interactions among the genes involved in cell growth and locomotion would perform better than relatively simple models such as ours, which focuses on two 'key' genes only. Secondly, our model contains only a genetically homogeneous population of tumor cells, which is advantageous if the focus is on phenotypic alternations such as in here. Yet, a more biologically realistic model will eventually need to incorporate genetic instability and thus heterogeneity.

Nonetheless, our goal here is to demonstrate that in this virtual brain tumor one can successfully combine the molecular-level gene expression dynamics with multicellular, macroscopic growth patterns using an agent-based modeling approach. As genomics data are becoming increasingly available for various tumor types our '*mol-micro-macro*' approach should provide a very helpful tool for investigating the crucial relationship between the phenotypic changes specific, microenvironmentally induced gene expression profiles cause on the one side and the expansion rate of the overall tumor system and its regional structural patterns on the other.

## ACKNOWLEDGEMENTS

This work has been supported by the Harvard-MIT (HST) Athinoula A. Martinos Center for Biomedical Imaging and the Department of Radiology at Massachusetts General Hospital. We thank Dr. Michael E. Berens (Translational Genomics Institute, Phoenix, AZ) for critical review of the manuscript.

# REFERENCES

[1] A. Giese, M.A. Loo, N. Tran, D. Haskett, S.W. Coons, and M.E. Berens, Int. J. Cancer **67**, 275 (1996a).

[2] L. Mariani, C. Beaudry, W.S. McDonough, D.B. Hoelzinger, T. Demuth, K.R. Ross, T. Berens, S.W. Coons, G. Watts, J.M. Trent, J.S. Wei, A. Giese, and M.E. Berens, J. Neurooncol. **53**, 161 (2001).

[3] Y. Mansury, M. Kimura, J. Lobo, and T.S. Deisboeck, J. theor. Biol. **219**, 343 (2002).

[4] Y. Mansury and T.S. Deisboeck, J. theor. Biol. In press.

[5] C. Herold-Mende, M.M. Mueller, M.M. Bonsanto, H.P. Schmitt, S. Kunze, and H.H. Steiner, Int. J. Cancer **98**, 362 (2002).

[6] D. Zagzag, B. Shiff, G.I. Jallo, M.A. Greco, C. Blanco, H. Cohen, J. Hukin, J.C. Allen, and D.R. Friedlander, Cancer Res. **62**, 2660 (2002).

[7] A. Giese, M.A. Loo, S.A. Norman, S. Treasurywala, and M.E. Berens, J Cell Sci. **109**, 2161. (1996b)

[8] P. Cavalla, and D. Schiffer, Anticancer Res. **17**, 4135 (1997).

[9] G. Karkavelas, S. Mavropoulou, G. Fountzilas, V. Christoforidou, A. Karavelis, G. Foroglou, and C. Papadimitriou, Anticancer Res. **15**, 531 (1995).

[10] L.J. Kirkegaard, P.B. DeRose, B. Yao, and C. Cohen, Am. J. Clin. Pathol, **109**, 69 (1998).

[11] Y. Mansury and T.S. Deisboeck, Physica A. In press.

[12] H. Bouterfa, A.R. Darlapp, E. Klein, T. Pietsch, K. Roosen, and J.C. Tonn, J. Neurooncol. **44**, 23 (1999).

[13] A. Lal, H. Peters, B. St Croix, Z.A. Haroon, M.W. Dewhirst, R.L. Strausberg, J.H. Kaanders, A.J. van der Kogel, and G.J. Riggins, J. Natl. Cancer Inst. **93**, 1337 (2001).

[14] J. Folkman, New Engl. J. Med. **285**, 1182 (1971).

[15] H.J. Bernsen, P.F. Rijken, H. Peters, J.A. Raleigh, J.W. Jeuken, P. Wesseling, A.J. van der Kogel, J. Neurosurg. **93**, 449 (2000).

[16] D.S. Steinbrech, B.J. Mehrara, P.B. Saadeh, G. Chin, M.E. Dudziak, R.P. Gerrets, G.K. Gittes, and M.T. Longaker, Plast. Reconstr. Surg. **104**, 738 (1999).

[17] P. Carmeliet and R.K. Jain, Nature **407**, 249 (2000).

[18] Y.C. Liu, W.L. Liu, S.T. Ding, H.M. Chen, and J.T. Chen, Exp Cell Res. **218**, 87 (1995).

[19] H.E. Stanley, L.A.N. Amaral, A.L. Goldberger, S. Havlin, P.Ch. Ivanov, and C.K. Peng, Physica A **270**, 309 (1999).

[20] C.-K. Peng, S.V. Buldyrev, S. Havlin, M. Simons, H.E. Stanley, and A.L. Goldberger, Phys. Rev. E **49**, 1685 (1994).

[21] S.V. Buldyrev, A.L. Goldberger, S. Havlin, R.N. Mantegna, M.E. Matsa, C.-K. Peng, M. Simons, and H.E. Stanley, Phys. Rev. E **51**, 5084 (1995).

[22] N.H. Packard, J.P. Crutchfield, J.D. Farmer, and R.S. Shaw, Phys. Rev. Lett. **45**, 712 (1980).

[23] J.D. Farmer and J.J. Sidorowich, Phys. Rev. Lett. **59**, 845 (1987)

[24] T. Hoshino and C.B. Wilson, J. Neurosurg. **42**, 123 (1975).

# FIGURE CAPTIONS

**FIG. 1.** Time Series of (a) Tenascin C and (b) PCNA gene expression for proliferating (*open triangles*) and migrating (*open rectangles*) tumor cells. Superimposed on the graphs are the time series of the average tumor diameter (*solid line*, *right* y-axis).

**FIG. 2.** Time series profile of *gTenascin* (*left* y-axis, *solid line*) and *gPCNA* (*right* y-axis, *dashed line*), averaged across all viable cells (i.e., including proliferating, migrating, and quiescent tumor cells).

**FIG. 3.** 2D snapshots of the tumor[a] during a typical simulation run at $\underline{t}$ = 308 (*left*) corresponding to the onset of the sharp rise in *gTenascin* and decline in *gPCNA* (see **Fig. 2**) and at $t$ = 371 (*right*) just before the first cell invades the peak of the second nutrient source.

[a]  Note the rapidly expanding, 'viable' tumor rim around a darkened, necrotic center.

**TABLE I.** cDNA Microarray data (of two experiments)**,** taken from [2].

**TABLE II. (1)** DFA results using the entire time series. **(2)** Correlation coefficients between gene expression and tumor diameter for the truncated series between the breakpoint (see **Fig. 2**) and the end of the simulation for (I) the entire tumor, (II) the NE region only, and (III) the SW region only. The numbers are averages over 20 simulation runs, with the standard deviations shown in brackets.

[a]    In addition to investigating the entire tumor we defined two regions of interest, i.e., the northeast (NE) quadrant (which hosts the second replenished nutrient source) and the southwest (SW) quadrant (which is farthest away from the second replenished nutrient source).

[b, c] Includes all viable tumor cells, i.e., proliferative, migrating and quiescent cells. The latter are defined as both non-proliferative and non-migratory.

# FIGURES AND TABLES

(a)

(b)

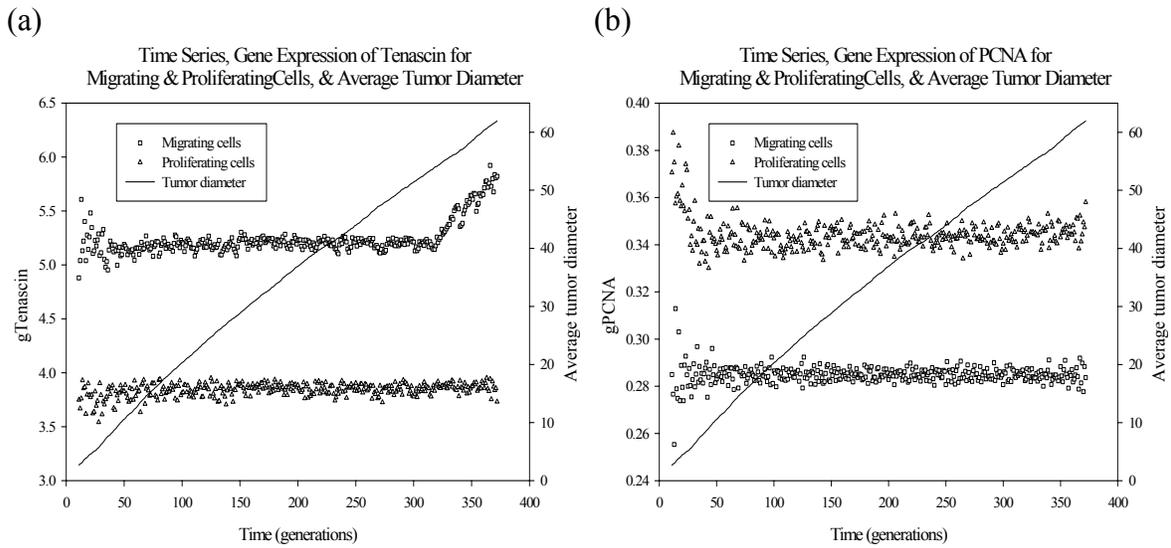

**FIG. 1**

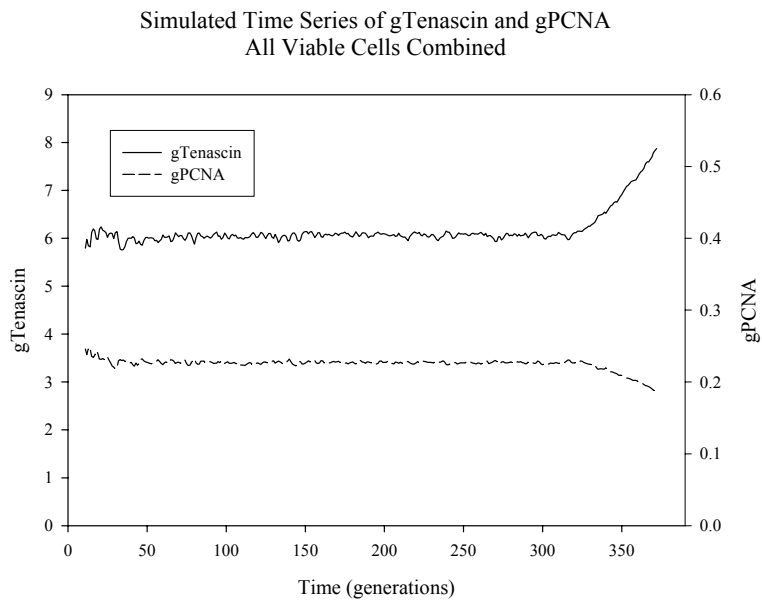

**FIG. 2**

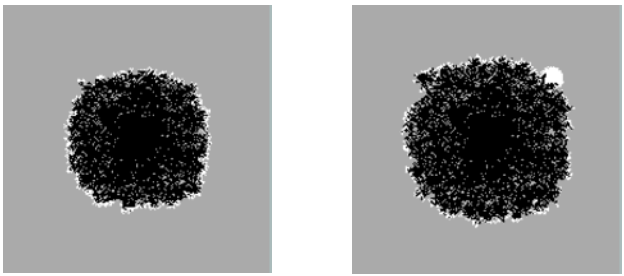

**FIG. 3**

| Clone ID | Clone Title | cDNA Array (1) [Ratio ECM/Plastic] | cDNA Array (2) [Ratio ECM/Plastic] | Mean |
|---|---|---|---|---|
| 23185 | Hexabrachion (Tenascin C) | 4.83 | 4.93 | 4.880 |
| 789182 | Proliferation nuclear cell antigen (PCNA) | 0.28 | 0.29 | 0.285 |

**TABLE I**

| | (I) Entire tumor | | (II) NE region[a] | | (III) SW region | |
|---|---|---|---|---|---|---|
| | **(1)** $\alpha$ | **(2)** $C_{T_B,T_E}$ | **(1)** $\alpha$ | **(2)** $C_{T_B,T_E}$ | **(1)** $\alpha$ | **(2)** $C_{T_B,T_E}$ |
| Tenascin, proliferating cells | 0.57 (0.01) | 0.16 (0.01) | 0.56 (0.01) | 0.35 (0.03) | 0.57 (0.01) | 0.29 (0.14) |
| Tenascin, migrating cells | 0.85 (0.03) | 0.95 (0.02) | 0.85 (0.06) | 0.95 (0.01) | 0.62 (0.02) | 0.54 (0.17) |
| Tenascin, all cells[b] | 1.32 (0.06) | 0.99 (0.01) | 1.16 (0.06) | 0.99 (0.00) | 0.63 (0.02) | 0.46 (0.08) |
| PCNA, proliferating cells | 0.63 (0.02) | 0.45 (0.02) | 0.58 (0.01) | 0.46 (0.02) | 0.58 (0.01) | 0.38 (0.29) |
| PCNA, migrating cells | 0.73 (0.03) | −0.93 (0.03) | 0.73 (0.04) | −0.93 (0.01) | 0.60 (0.02) | −0.32 (0.17) |
| PCNA, all cells[c] | 1.06 (0.04) | −0.99 (0.00) | 0.93 (0.04) | −0.99 (0.01) | 0.60 (0.01) | −0.18 (0.04) |

**TABLE II**